\documentstyle[12pt,psfig]{article}

\font\blackboard=msbm10 at 12pt
\font\blackboards=msbm7
\font\blackboardss=msbm5
\newfam\black
\textfont\black=\blackboard
\scriptfont\black=\blackboards
\scriptscriptfont\black=\blackboardss
\def\bb#1{{\fam\black\relax#1}}
\def\BZ{\bb Z}

\def\Re{{\rm Re\ }}
\def\Im{{\rm Im\ }}

\newcommand{\NP}{{\em Nucl.\ Phys.\ }}
\newcommand{\PL}{{\em Phys.\ Lett.\ }}

\newcommand{\CMP}{{\em Comm.\ Math.\ Phys.\ }}
\newcommand{\MPL}{{\em Mod.\ Phys.\ Lett.\ }}
\newcommand{\PRL}{{\em Phys.\ Rev.\ Lett.\ }}

\newcommand{\tr}{{\rm Tr}}

\newcommand{\lag}{{\cal L}}
\newcommand{\newcaption}[1]{\centerline{\parbox{6in}{\caption{#1}}}}

\newcommand{\inn}{\!\cdot\!}

\newcommand{\gone}[1]{}
\begin{document}
\pagestyle{plain}
\setcounter{page}{1}

\baselineskip16pt

\begin{titlepage}

\begin{flushright}
PUPT-1693\\
hep-th/9703217
\end{flushright}
\vspace{20 mm}

\begin{center}
{\Large \bf Fluctuation Spectra of Tilted and Intersecting D-branes
from the Born-Infeld Action}

\vspace{3mm}

\end{center}

\vspace{10 mm}

\begin{center}
{Akikazu Hashimoto and Washington Taylor IV}

\vspace{2mm}

{\small \sl Department of Physics} \\
{\small \sl Joseph Henry Laboratories} \\
{\small \sl Princeton University} \\
{\small \sl Princeton, New Jersey 08544, U.S.A.} \\
{\small \tt aki, wati @princeton.edu}

\end{center}

\vspace{1cm}

\begin{abstract}
We consider the spectra of excitations around diagonal and
intersecting D-brane configurations on tori.  These configurations are
described by constant curvature connections in a dual gauge theory
description.  The low-energy string fluctuation spectrum is reproduced
exactly by the gauge theory in the case of vanishing field strength;
however, this correspondence breaks down for fixed nonzero field
strength.  We show that in many cases the full Born-Infeld action
correctly captures the low-energy spectrum in the case of
non-vanishing field strength.  This gives a field theory description
of the low-energy physics of systems of diagonally wound branes and
branes at angles as considered by Berkooz, Douglas and Leigh.  This
description extends naturally to non-supersymmetric configurations,
where the tachyonic instability associated with brane-anti-brane
systems appears as an instability around a saddle point solution of
the corresponding Yang-Mills/Born-Infeld theory.  In some cases, the
field theory description requires a non-abelian generalization of the
Born-Infeld action.  We follow Tseytlin's recent proposal for
formulating such an action.  In the case of intersecting branes, the
non-abelian Born-Infeld theory produces a transcendental relation
which comes tantalizingly close to reproducing the correct spectrum;
however, a discrepancy remains which indicates that a further
clarification of the non-abelian Born-Infeld action may be necessary.
\end{abstract}

\vspace{1cm}
\begin{flushleft}
March 1997
\end{flushleft}
\end{titlepage}
\newpage

\section{Introduction}
\label{Intro}

D-branes \cite{polchinski} have recently emerged as a key ingredient
driving the non-perturbative dynamics of string theory.  These objects
provide an exact description of string theory solitons, and have led
to remarkable developments towards understanding the nature of stringy
black holes \cite{Strominger:1996,cm}.  D-branes also provide a
fascinating connection between string theory and gauge theories. The
low-energy effective dynamics of $N$ parallel D-branes is precisely
that of supersymmetric Yang-Mills theory with gauge group $U(N)$
\cite{ed}; such a low-energy effective theory has in turn been
conjectured to provide a reformulation of string theory itself
\cite{BFSS}.

The goal of this paper is to extend further the range of phenomena in
D-brane physics which can be described precisely in the language of
Yang-Mills field theory or its extension to Born-Infeld theory.  In
particular, we consider the correspondence between the perturbative
fluctuation spectra of D-brane configurations wrapped on a torus and
gauge theories with a constant background field on the dual
torus\footnote{Such a correspondence at the level of the vacuum state
was considered previously using the boundary state formalism in
\cite{li,ck}.}.  Constant background field configurations in
Yang-Mills theory on the torus are T-dual to D-branes which are either
diagonally wrapped (``tilted'') \cite{Polchinski:TASI} or intersecting
at an angle \cite{BDL96}.  Related recent work on branes at an angle
can be found in
\cite{SanjayZack,vijayrob97,Breckenridge:1997,Gauntlett:1997,Behrndt:1997,Costa:1997,Hambli:1997}.
The fluctuation spectra of gauge theories in a constant background
field have been investigated previously in \cite{vanBaal84}.  We
compare these results to what is expected from the D-brane side.  The
constant background field strength in the Yang-Mills theory gives rise
to a new scale in the problem in addition to the string scale and the
radius of the torus.  In an appropriate scaling limit, we show that
Yang-Mills theory is insufficient to reproduce the spectrum of open
strings on the D-branes and must be replaced by the full Born-Infeld
action.  We find that in many cases the full Born-Infeld action
reproduces in a field theory context the exact spectrum of low-energy
string excitations around the D-brane background.

The story is slightly more subtle in the non-abelian case.  The
excitation spectra for intersecting D-branes and the corresponding
dual gauge fields were discussed in \cite{BDL96} and \cite{vanBaal84}
respectively.  However, these spectra agree only to lowest order in
the angle of intersection.  We investigate the resolution of this
discrepancy by including higher order terms in the non-abelian
Born-Infeld action.  Although the non-abelian generalization of the
Born-Infeld action has yet to be understood in full, a formulation
involving a symmetrized trace was proposed recently by Tseytlin
\cite{NDBI}.  We find that this prescription for non-abelian
Born-Infeld fails to reproduce the exact fluctuation spectrum seen in
string theory.  Nonetheless, we encounter remarkable relations encoded
in the general structure of Tseytlin's non-abelian Born-Infeld action
which provide a hint indicating how this discrepancy might ultimately
be resolved.  Since the relation which emerges from the analysis of
the action is highly non-trivial, we believe that there is a strong
element of truth in Tseytlin's proposal.  However, we do not
completely resolve the discrepancy at finite angles, and we leave this
as a problem for future investigations.

A particularly interesting class of systems of the type considered
here are those corresponding to non-supersymmetric D-brane
configurations.  On $T^4$ constant gauge field backgrounds are stable
if and only if the gauge field is (anti)-self-dual.  In \cite{BDL96}
it was argued that the (anti)-self-dual condition is equivalent to
the condition that the dual D-brane configuration preserve enough
supersymmetry to be a BPS state.  Considering non-supersymmetric
D-brane configurations, we find that the tachyonic instability of the
string ground state appears naturally in the field theory language as
an instability of a saddle point configuration in the Yang-Mills/Born-Infeld
theory.  This provides a natural framework for studying aspects of
brane-anti-brane interactions in the context of  field theory.

The organization of this paper is as follows: In section 2 we
consider D-brane configurations which are tilted with respect to the
torus; these configurations correspond to gauge theory backgrounds
with constant field strengths in the central $U(1)$ part of $U(N)$.
The necessity for replacing the Yang-Mills action with the Born-Infeld
action in order to exactly reproduce the fluctuation spectrum emerges
from this analysis.  We also discuss the issue of fractional
quantization from the gauge theory point of view in this section.  In
section 3 we consider systems of intersecting D-branes which
correspond to reducible connections in the dual gauge theory.  We
discuss the discrepancy between the spectra and its possible resolution by the
non-abelian Born-Infeld action.  We also discuss here the
D-brane/Born-Infeld correspondence for non-supersymmetric D-brane
configurations and the associated tachyonic instability.

\section{Tilted D-branes and the Born-Infeld Action}
\label{sec:tilted}

\subsection{Review of D-branes and T-duality}

We begin with a brief review of some salient features of D-branes and
their behavior under T-duality; for more background see
\cite{Polchinski:TASI}.  

In this paper we are concerned with D-branes in type II string theory.
In type IIA (IIB) string theory, D-branes with even (odd) dimension
appear.  When the theory is compactified on a torus $T^d$ there is a
T-duality symmetry in each compactified direction which exchanges
Neumann and Dirichlet string boundary conditions and which therefore
changes the dimension of a D-brane by one.  The dynamics of a single
D-brane are controlled by the Born-Infeld  action,
which reduces in the case of small field strength and flat background
metric to a supersymmetric $U(1)$ Yang-Mills action with adjoint
scalars corresponding to transverse fluctuations \cite{DBI}.
When $N$ D-branes become coincident, the Yang-Mills action is extended
to a supersymmetric non-abelian 
$U(N)$ Yang-Mills theory \cite{ed}.  The
generalization of the full Born-Infeld action to the non-abelian case
has not yet been completely understood; however, recent progress in
this direction was made by Tseytlin \cite{NDBI}.

When D-branes are living in a space which has been compactified in a
transverse direction, there are winding strings which wrap around the
compactified direction an arbitrary number of times.  Under T-duality,
string winding and string momentum are exchanged so that these winding
strings naturally become momentum modes of the gauge field under
duality in the compact direction.  At the level of the 
Yang-Mills action, this duality can be made manifest.  As discussed in
\cite{Taylor:1996,GRT:1996}, the winding strings can be packaged in an
infinite matrix which corresponds to the operator $2 \pi \alpha' (i
\partial + A)$ in the dual gauge theory.  Under this correspondence,
the fluctuation spectrum of strings on a D-brane which is unwrapped in
some set of compact dimensions is precisely captured by the dual gauge
theory.

When we consider D-branes which are not parallel to the generators of
a perpendicular torus, however, or systems of intersecting D-branes,
we must be more careful.  It was pointed out in \cite{witinst,doug}
that nontrivial gauge configurations can carry D-brane charge.  This
follows from the Chern-Simons term in the D-brane action which
couples higher powers of the gauge field strength $F$ to RR fields.
When we have a system of tilted or intersecting D-branes on a torus,
T-dualizing to a gauge theory on the torus will generally lead to a
topologically nontrivial bundle whose Chern classes determine the
D-brane charges on various subtori.  Such bundles were discussed in
the $SU(N)$ case in  \cite{tHooft81}, and in the context of D-branes
in \cite{SanjayZack}.  The approach we will take in this
paper is to consider fixed configurations of flat D-branes on the
torus which may be tilted or intersecting.  These D-brane
configurations map under T-duality to constant field strength
backgrounds in the dual gauge theory.  We will compare the fluctuation
spectra of the D-brane configurations as calculated in string theory
to the fluctuation spectra of the gauge fields around the constant
background.  As we shall see, in many situations the action must be
generalized to the full non-abelian Born-Infeld  action to achieve an exact
correspondence between these two spectra.


\subsection{1-brane on $T^2$ with winding $(1, q)$}

Let us begin with the simplest example of a gauge field configuration
corresponding to the T-dual of a tilted D-brane. Consider a 2-brane
wrapped on $T^2$ with periods $L_1$ and $L_2$. The fluctuation
spectrum of this theory is precisely that of SUSY
$U(1)$ Yang-Mills theory in
2+1 dimensions around a vanishing background field.  There are gauge
fields corresponding to the regular bosonic $U(1)$ theory, adjoint
scalars corresponding to transverse fluctuations, and various
fermionic fields.  Throughout this paper we will restrict attention to
the fluctuation spectra of the bosonic modes, although a similar
analysis could easily be performed for the fermion modes.  Consider
turning on a background
\begin{equation}
\begin{array}{lll}
A^0_1 & = & 0  \\
A^0_2 & = & F_{0} x_1.
\end{array}
\label{background}
\end{equation}
The flux quantization condition requires that $F_{0}$ be an integer
multiple of the flux quantum:
$$F^0_{}
=F^0_{21} = \frac{2 \pi}{L_1 L_2} q,\qquad q \in  \BZ.$$
Such a configuration corresponds to a $U(1)$ bundle with first Chern
class $C_1 = q$.  For concreteness, let us take $q=1$. From the point
of view of $U(1)$ gauge theory, it is straightforward to compute the
spectrum of fluctuations around this background.  In terms of the
fluctuation $\delta A_\mu$ defined through
$$A_\mu=A_\mu^0+\delta A_\mu,$$
the Lagrangian is simply (up to an overall constant)
$$\lag = F_{\mu \nu} F^{\mu \nu} = F^0_{\mu \nu} F^{0\, \mu
\nu} +2F^0_{\mu \nu} (\partial^\mu \delta A^\nu - \partial^\nu \delta
A^\mu) + (\partial_\mu \delta A_\nu - \partial_\nu \delta A_\mu)
(\partial^\mu \delta A^\nu - \partial^\nu \delta A^\mu)
 $$
We can ignore the additive constant and the total derivative terms, so
one recovers the original Yang-Mills Lagrangian
\begin{equation}
\lag = (\partial_\mu \delta A_\nu - \partial_\nu \delta A_\mu)
(\partial^\mu \delta A^\nu - \partial^\nu \delta A^\mu)
\label{original}
\end{equation}
whose physical spectrum is simply that of the quantized momenta
\begin{equation}
E^2 = \left(\frac{2\pi}{L_1}n_1\right)^2 +
\left(\frac{2\pi}{L_2}n_2\right)^2.
\label{gaugespec}
\end{equation}
The spectrum for the transverse bosonic fields is clearly the same, as
they obey the same periodic boundary conditions.

\begin{figure}
\psfig{file=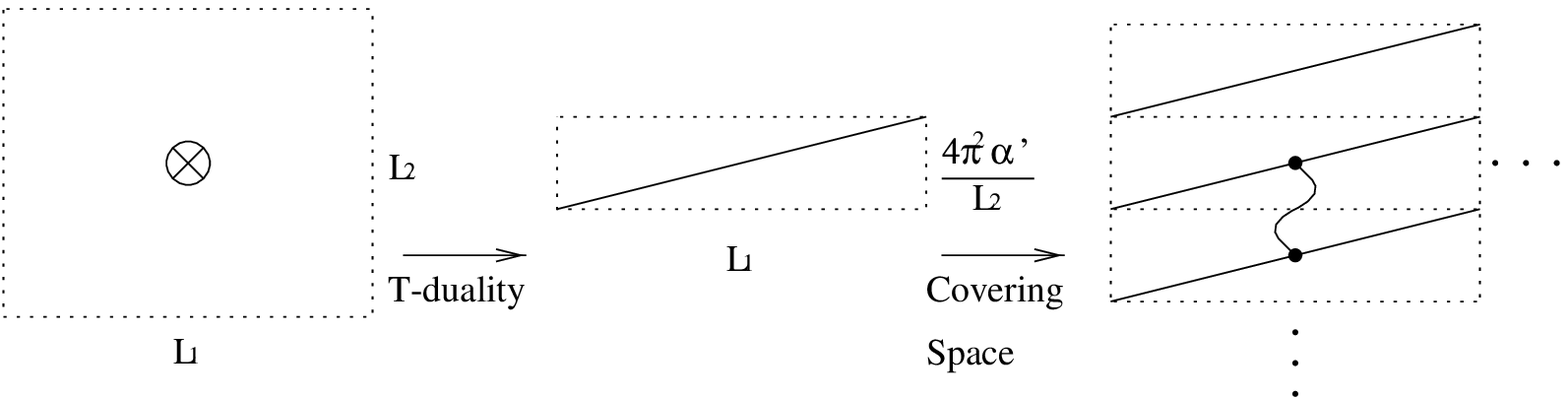}
\newcaption{A 2-brane with unit flux quantum and its T-dual \label{tilt1}}
\end{figure}

Let us now consider the D-brane configuration given after
T-duality in the $x_2$ direction.  In the gauge theory configuration
described above, we have a single unit of 2-brane charge corresponding
to the $U(1)$ gauge group.  The flux $F_{0}$ carries a single unit of
0-brane charge.  Under T-duality, we get a D-brane configuration with
one unit of 1-brane charge in each direction $x_1, x_2$.  T-duality
inverts the radius of the torus along the $x_2$ direction, so the
dimensions of the dual torus are $L_1 \times 4 \pi^2 \alpha'/L_2$.
The background field given by (\ref{background}) can be interpreted
under T-duality as the transverse coordinate of a 1-brane on $x_1$,
$$X_2 =  2 \pi \alpha' A^0_2 = \frac{4 \pi^2 \alpha' }{L_1 L_2} x_1.$$
This shows that the 1-brane is wound diagonally  on the dual
torus, as expected of a (1, 1) bound state (see figure \ref{tilt1}).
The field theory limit of this configuration corresponds to taking
$L_1$ and $L_2$ to be of the same order of magnitude while sending
$\alpha'/L_1^2$ to zero. In this limit, the dual torus becomes long and
thin. The low-lying states in the T-dual description are therefore
given by momentum modes along the 1-brane and winding modes in the
direction perpendicular to the 1-brane.  The winding modes can also be
thought of as strings stretching between adjacent 1-branes in the
covering space of the dual torus.  From the point of view of string
theory in a background given by such a D-brane configuration, we
expect the spectrum of low-lying excitations to be given by
\begin{equation}
E^2 = \left(\frac{2\pi}{L'_1}n_1\right)^2 +
\left(\frac{L'_2}{2\pi \alpha' }n_2\right)^2,
\label{tmpspec}
\end{equation}
where $L'_1$ is the length of the 1-brane along the diagonal
$$L'_1  = L_1 \sqrt{1 + \left(\frac{4 \pi^2 \alpha'}{L_1
L_2}\right)^2} = L_1 \sqrt{1+(2 \pi \alpha' F_{0})^2} $$ 
and $L'_2$ is the distance between adjacent 1-branes
$$L'_2 = \frac{1}{\sqrt{1 + \left(\frac{4 \pi^2 \alpha'}{L_1
L_2}\right)^2}} \left(\frac{4 \pi^2 \alpha'}{L_2}\right) 
=\frac{1}{\sqrt{1 + (2 \pi \alpha' F_{0})^2}} \left(\frac{4 \pi^2
\alpha'}{L_2}\right).$$ 
In terms of the original variables, (\ref{tmpspec}) becomes
\begin{equation}
E^2 = \frac{1}{1+(2 \pi \alpha'
F_{0})^2}\left(\left(\frac{2\pi}{L_1}n_1\right)^2 + 
\left(\frac{2\pi}{L_2}n_2\right)^2\right) \label{dbranespec}
\end{equation}
Comparison of eq. (\ref{gaugespec}) and (\ref{dbranespec})
indicates that they disagree by a factor of $(1+(2 \pi\alpha'
F_{0})^2)$.  We will now show that the spectrum of excited states
arising from the Born-Infeld action in 2+1 dimensions gives rise to the
correct factor of $(1+(2 \pi \alpha' F_{0})^2)$, correcting
(\ref{gaugespec})
to precisely match
the string theory result found in equation (\ref{dbranespec}).  
The Born-Infeld
action in 2+1 dimensions is given by (up to an overall constant)
\begin{equation}
S = \int d^3 x \sqrt{-\det (\eta_{\mu \nu} + 2 \pi \alpha' F_{\mu \nu}) }
\label{Born-Infeld}
\end{equation}
where $\eta_{\mu \nu}$ is the Minkowski metric ${\rm Diag}\{-1,1,1\}$,
which will henceforth be used to raise and lower the Lorentz indices.
Just as in the case of the original Yang-Mills theory, the dynamics of
small fluctuations is described by expanding the action around a fixed
background. Substituting
$$A_\mu = A^0_\mu + \delta A_\mu$$
into (\ref{Born-Infeld}) and expanding to quadratic order in $\delta A_\mu$,
one finds that up to an overall scaling factor the quadratic term in
the action is given by
\begin{equation}
\lag_2 = 
 \sqrt{1+(2 \pi \alpha' F_{0})^2} 
\left(
(1+2 \pi \alpha' F_0)^{-1}_{\mu \nu} \tilde{F}^{\nu \lambda} (1+2 \pi \alpha' F_0)^{-1}_{\lambda \sigma}
 \tilde{F}^{\sigma 
\mu}\rule{0ex}{3ex}- 
\frac{1}{2} \left((1+2 \pi \alpha' F_0)^{-1}_{\mu \nu} \tilde{F}^{\nu \mu}\right)^2
\right)
\label{fluctBorn-Infeld}
\end{equation}
where
$$\tilde{F}_{\mu \nu} = \partial_\mu \delta A_\nu - \partial_\nu
\delta A_\mu$$ 
is the fluctuation.  It will be convenient to separate $(1+2 \pi \alpha' F_0)^{-1}$ 
into its symmetric and antisymmetric components:
$$(1+2 \pi \alpha' F_0)^{-1}_{\mu \nu} = g_{\mu \nu} + B_{\mu \nu}.$$
Using this decomposition and ignoring the overall factor of
$\sqrt{1+(2 \pi \alpha' F_{0})^2}$,  the action takes the form
$$
\lag_2 = g^{\mu \nu} \tilde{F}_{\nu \lambda} g^{\lambda \sigma}
\tilde{F}_{\sigma \mu} 
+\left( B^{\mu \nu} \tilde{F}_{\nu \lambda} B^{\lambda \sigma}
\tilde{F}_{\sigma \mu} - \frac{1}{2} \left(B^{\mu \nu} \tilde{F}_{\mu
\nu} \right)^2 \right). 
$$
Further computation shows that the terms containing $B^{\mu \nu}$
vanish, and $g^{\mu \nu}$ is a diagonal matrix $ g^{\mu\nu} = {\rm
Diag} \{-1,g_{(12)},g_{(12)} \}$ with
$$g_{(12)}= \frac{1}{1+(2 \pi \alpha' F_{0})^2}.$$
Compared to the dynamics of small fluctuations in Yang-Mills theory
given by equation (\ref{original}), the Born-Infeld action has effectively
rescaled the spatial coordinates by a factor of $\sqrt{1+(2 \pi
\alpha' F_{0})^2}$. This properly rescales the spectrum of the
momentum modes so that there is a precise agreement with the spectrum
of low-lying open string states (\ref{dbranespec}) seen from the
D-brane point of view.

\begin{figure}
\psfig{file=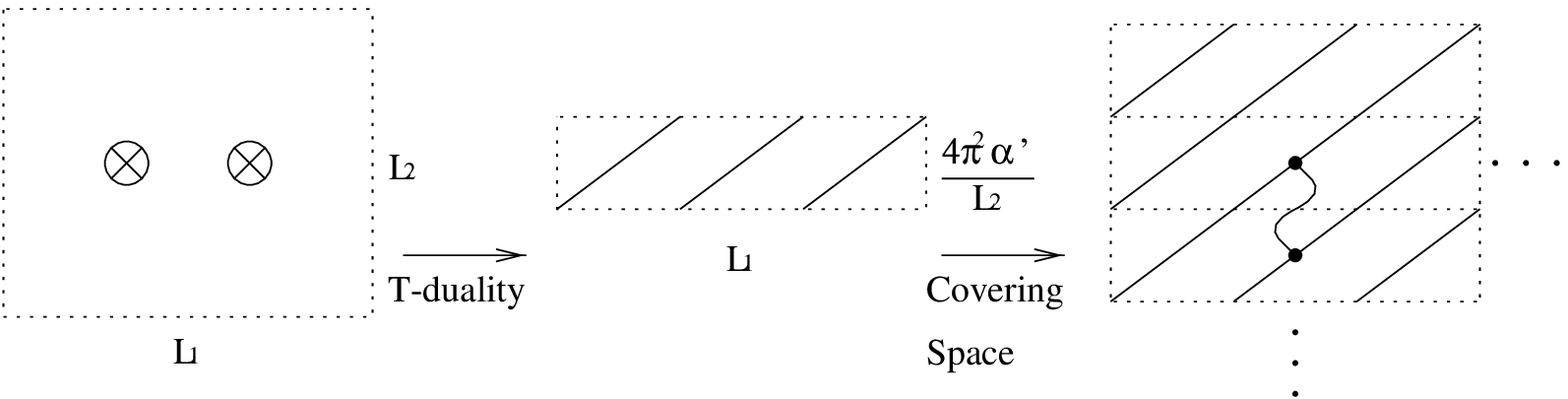}
\newcaption{A 2-brane with $q$ flux quanta and its T-dual
\label{tilt2}}
\end{figure}

This exercise of establishing the necessity for considering the Born-Infeld
action over the Yang-Mills action in order to exactly match the
string theory spectrum might seem purely academic in light of the fact
that in the field theory limit $2 \pi \alpha' F_{0}$ goes to zero
and the factor in question, $1+ (2 \pi \alpha' F_{0})^2$, is
identically 1 to leading order in $\alpha'/L_1^2$. However, there is a
way to take a field theory limit keeping $2 \pi \alpha' F_{0}$
constant. Consider the same background as before
$$F_0 = \frac{2 \pi}{L_1 L_2} q,\qquad q \in \BZ.$$
with $q$ not equal to 1.  From the D-brane point of view, this corresponds to 
a 1-brane with wrapping number $(1,q)$.  (See figure \ref{tilt2}.)
Again, we expect the spectrum to be given by (\ref{tmpspec}), and just as
before, we obtain
\begin{equation}
E^2 = \frac{1}{1+(2 \pi \alpha'
F_{0})^2}\left(\left(\frac{2\pi}{L_1}n_1\right)^2 + 
\left(\frac{2\pi}{L_2}n_2\right)^2\right) 
\end{equation}
Now, we can take the field theory limit by keeping $L_1$, $L_2$ and $q
\alpha'/L_1 L_2$ constant and sending $q$ to infinity. This time, the
difference between Born-Infeld and Yang-Mills theory is non-trivial even at leading
order in $\alpha'/L_1^2$.

\begin{figure}
\centerline{\psfig{file=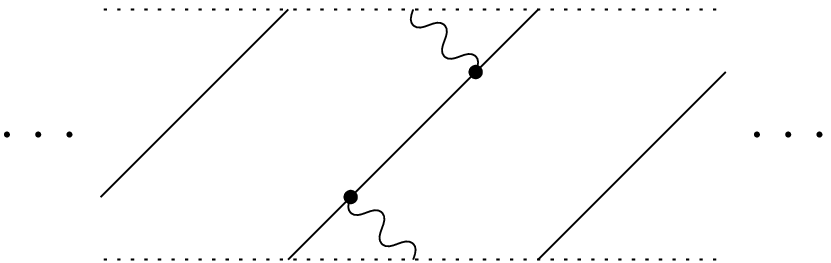}}
\newcaption{Winding modes of open strings attached to a tilted D-brane
are not local from the world-volume observer's point of view.
\label{nonlocal}}
\end{figure}

The fact that the Born-Infeld action captures the dual string spectrum
more correctly than Yang-Mills theory is very natural in light of the
fact that the dynamics on the world-volume of a D-brane is indeed
described by the Born-Infeld action and Yang-Mills theory is only its
leading approximation in the field theory limit for a fixed
background.  In fact, the same type of system as the one we have
discussed in this section was analyzed in
\cite{Fradkin-Tseytlin,acny,Argyres-Nappi} from the point 
of view of string theory in a background gauge field, giving essentially
the same results.  The interesting point about this type of
configuration is that by scaling the background as above it is
possible to retain the structure of Born-Infeld dynamics even in the
field theory limit.  The extra terms arising from the Born-Infeld
action are usually subleading in the $\alpha'$ expansion and are
associated with the non-locality of string theory.  We can in fact see
that the winding strings of minimum length attached to the tilted
branes are not local objects from the world-volume observer's point of
view, as is illustrated in figure \ref{nonlocal}.
The results from this section indicate that it is possible to
enhance the non-locality of string theory even in the field theory
limit by scaling up the background field strength. In the presence of
such a background, the interaction terms in the Born-Infeld action will induce
a mass shift in the spectrum of excited states to precisely reproduce
the string theory spectrum.


\subsection{1-brane on $T^2$ with winding $(p, q)$ \label{sec:pq}}

In the previous subsection we considered only configurations with a
single 2-brane in the gauge theory picture, and all our analysis was
done in the context of the abelian Born-Infeld action.  In this subsection we
discuss a more general configuration which is T-dual to a D1-brane on
$T^2$ with wrapping numbers $(p,q)$ with $p, q$ relatively prime.  To
describe such a configuration we need to define a $U(p)$ bundle over
$T^2$ whose connections have total flux $2 \pi q/p$.  The flux is
quantized in units of $2 \pi/p$ since the gauge group is $U (p)= (U(1)
\times SU(p))/{\BZ}_p$.  We can describe a $U(p)$ bundle in the
fundamental representation by giving the boundary overlap functions
for sections
\begin{eqnarray*}
\phi (x_1 + L_1, x_2) & = & \Omega_1 (x_2)\phi (x_1, x_2) \\
\phi (x_1, x_2 + L_2) & = & \Omega_2 (x_1)\phi (x_1, x_2).
\end{eqnarray*}
The condition that these boundary conditions give a well-defined
$U(p)$ bundle is
\begin{equation}
\Omega_2^{-1} (L_1) \Omega_1^{-1} (0) \Omega_2 (0) \Omega_1 (L_2) = 1.
\label{eq:bundlecondition}
\end{equation}
To have a total flux of $2 \pi q/p$ we can decompose these boundary
conditions into abelian and non-abelian components.  The abelian
component should have flux $2 \pi q/p$, while the non-abelian component
should have an 't Hooft \cite{tHooft81} twist giving a flux $2 \pi
\tilde{q}/p$ where $0 \leq \tilde{q} < p$ and $\tilde{q} \equiv q
({\rm mod}\; p)$.  A standard choice of gauge for defining a bundle of
this type is to pick 
\begin{eqnarray*}
\Omega_1 (x_2) & = &  e^{2 \pi i (x_2/ L_2) (q/p) I}U^q\\
\Omega_2 (x_1) & = &  V
\end{eqnarray*}
where $U$ and $V$ are the diagonal and shift matrices
\begin{equation}
U = 
\pmatrix{
1& & & \cr
& e^{2\pi i\over p} & & \cr
& & \ddots & \cr
& & & e^{2\pi i (p-1)\over p}
},\qquad
V =
\pmatrix{
 & 1 & & \cr
 &   & 1 & \cr
 &   &   & \ddots \cr
1&   &   &  
} 
\end{equation}
Choosing this set of boundary conditions, however, complicates the
discussion of T-duality in the $x_2$ direction.
We will therefore instead choose the boundary conditions
\begin{eqnarray*}
\Omega_1 (x_2) & = & e^{2 \pi i (x_2/L_2) T}V^q\\
\Omega_2 (x_1) & = & 1
\end{eqnarray*}
with $T$  being the diagonal matrix
\begin{eqnarray*}
T & = &  (q/p) I +
{\rm Diag} (-\tilde{q}/p, \ldots, -\tilde{q}/p,
1-\tilde{q}/p, \ldots, 1-\tilde{q}/p)\\
 & = &  {\rm Diag} (n, \ldots, n, n + 1, \ldots, n + 1)
\end{eqnarray*}
where $n$ is the integral part of $q/p$, and where the multiplicities
of the diagonal elements of $T$ are $p-q$ and $q$ respectively.  Since
$T$ has integral diagonal elements, we have 
\[
\Omega_1 (L_2) = V^q = \Omega_1 (0)
\]
so
(\ref{eq:bundlecondition}) is clearly satisfied.  The abelian part of
$T$ is just $(q/p) I$ so this bundle has the correct flux.
Since the boundary conditions are trivial in the $x_2$ direction, we
can discuss T-duality in this direction without encountering undue
complications.

\begin{figure}
\psfig{file=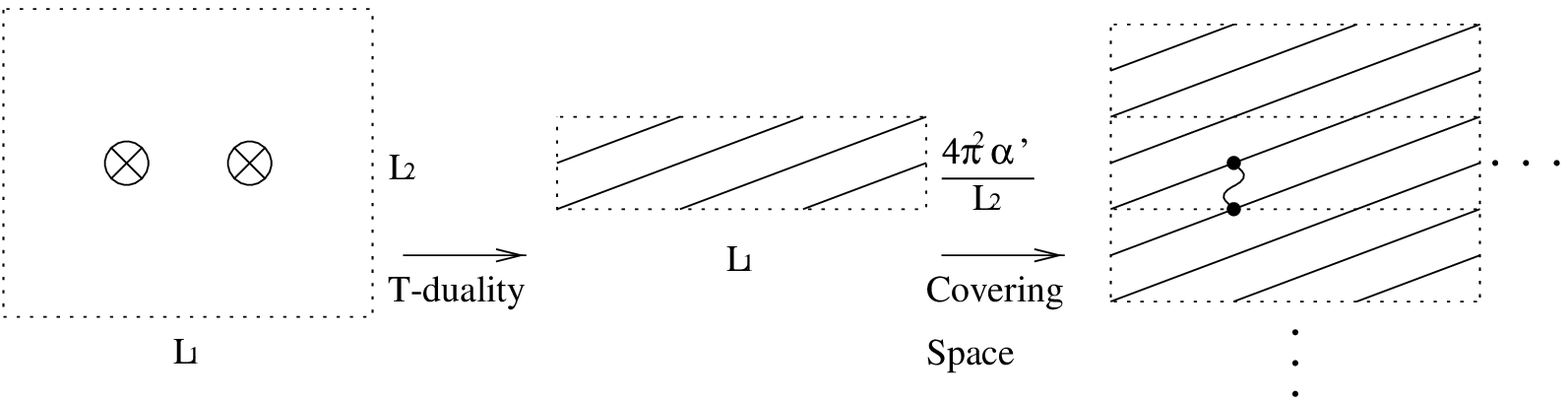}
\caption{$p$ 2-branes with $q$ flux quanta and the T-dual  $(p, q)$ 1-brane
\label{tilt3}}
\end{figure}

The gauge fields $A_\mu$ on this bundle will satisfy the
boundary conditions
\begin{eqnarray}
A_1 (x_1 + L_1, x_2) & = &  e^{2 \pi i (x_2/L_2) T} 
V^q A_1 (x_1, x_2) V^{-q} e^{-2 \pi i (x_2/L_2) T} \nonumber \\
A_1 (x_1, x_2 + L_2) & = &    A_1 (x_1, x_2) \nonumber \\
A_2 (x_1 + L_1, x_2) & = &  e^{2 \pi i (x_2/L_2) T} V^q A_2 (x_1, x_2) 
V^{-q} e^{-2 \pi i (x_2/L_2) T}
  + \left(\frac{2 \pi}{L_2 }\right) T  \nonumber \\
A_2 (x_1, x_2 + L_2) & = &    A_2 (x_1, x_2) \label{bcond}
\end{eqnarray}
The adjoint scalar fields $X^i$ will satisfy the same boundary
conditions as $A_1$.
The constant curvature background corresponding to these boundary
conditions is
\begin{eqnarray*}
A^0_1  & = & 0 \\
A^0_2  & = & F_0^{} x_1 I +  \frac{2 \pi}{L_2} 
{\rm Diag} (0,  1/p, \ldots, (p -1)/p) 
\end{eqnarray*}
where
$$F_{0} = \frac{2 \pi}{L_1 L_2} \frac{q}{p}.$$
The T-dual along $x_2$ gives rise to the type of D1-brane
configuration illustrated in figure \ref{tilt3}, with
\[
X_2   =  2 \pi \alpha' F_{0} x_1 +  \frac{4 \pi^2 \alpha'}{L_2} 
{\rm Diag} (0,  1/p, \ldots, (p -1)/p) 
\]
It is straightforward to derive the spectrum of open strings around
this D-brane background:
\begin{equation}
E^2 = \frac{1}{p^2+\left(\frac{4 \pi^2 \alpha'}{L_1 L_2} \right)^2
q^2}\left(\left(\frac{2\pi}{L_1}n_1\right)^2 + 
\left(\frac{2\pi}{L_2}n_2\right)^2\right).
\end{equation}
In order to reproduce this spectrum in field theory, we need the
non-abelian generalization of the Born-Infeld action.  Although there are some
ambiguities in defining the non-abelian Born-Infeld action, a procedure for
defining this action was recently proposed by Tseytlin in
\cite{NDBI}. In the following, we will simply follow this proposal,
which gives an action of the form 
$$\lag = {\rm STr} \sqrt{-\det(\eta_{\mu \nu} + 2 \pi\alpha' F_{\mu \nu})}$$
where ``STr'' indicates that the trace is to be taken after we
symmetrize all the non-commuting products.  Our analysis is simplified
drastically by the fact that the background gauge field is
proportional to the identity and commutes with everything.  Expanding
the non-abelian Born-Infeld action to second order will therefore lead
to
\begin{equation}
\lag_2 = \sqrt{1+(2 \pi \alpha' F_{0})^2} \; \tr[ g^{\mu \nu}
\tilde{F}_{\nu \lambda} g^{\lambda \sigma} \tilde{F}_{\sigma \mu}]
\label{abelian}
 \end{equation}
with the same $g^{\mu \nu}$ we encountered previously.
%
%
%
Just as in the previous case, we see that the effect of including the
correction arising from the full Born-Infeld type interaction is the
rescaling of the spectrum by a factor of $1 + (2 \pi \alpha'
F_{0})^2$.  Thus, we have a spectrum given by
\begin{equation}
E^2 = \frac{1}{1+ (2\pi\alpha' F_{0})^2}
\left(k_1^2 + k_2^2 \right).
\label{back}
\end{equation}
where $k_i$ is the eigenvalue of the covariant derivative operator
$D_i$ acting on the modes satisfying the boundary conditions
(\ref{bcond}).  Due to the twisting in the boundary conditions, the
quantization conditions of these modes are complicated slightly, and
the $p^2$-fold degeneracy corresponding to the number of independent
components of matrices in the adjoint of $U(N)$ is lifted.  Let us now
consider the effects of the boundary conditions on the periodicity of
the various modes.  The inhomogeneous term in the boundary condition
for $A_2$ does not affect the periodicity structure of the fluctuation
modes since it is taken care of by the background field.  Thus, the
fluctuation spectrum will be the same for each of the gauge fields and
the transverse scalars.  Each of these fields is described by a matrix
containing $p^2$ elements.

It is straightforward to show that the homogeneous part of the boundary
condition (\ref{bcond}) is satisfied by modes given by
$$
\delta A(x_1,x_2)  =  \Lambda 
e^{2 \pi i (m_1 x_1/L_1  + m_2 x_2/L_2)}
$$
where $\Lambda$ is a $p\times p$ matrix
$$
\Lambda = 
{\rm Diag}\{1, \omega, \omega^2, \ldots, \omega^{p-1} \} \inn
\left(\begin{array}{rclrcl}
&&&    e^{-2 \pi i x_2/L_2} \!\!\!\!\!\!\!\!\!\!\!\!\!\!\!\!\!\!\!\! & & \\
&&&    & \ddots  & \\
&&&    & &  e^{-2 \pi i x_2/L_2}\\
    1 & & &&& \\
    & \ddots  & &&& \\
    & & 1 &&&
\end{array}\right)
\begin{array}{l}
\left.\rule{0ex}{5.5ex}\right\}r\\
\left.\rule{0ex}{5.5ex}\right\} p-r
\end{array}
$$
where $\omega = e^{2 \pi i m_1 s}$ for $qs \equiv 1\ 
({\rm mod}\; p)$.  The
 boundary condition (\ref{bcond}) further implies that
$$A(x_1+p L_1, x_2) = A(x_1,x_2)$$
indicating that $m_1$ takes on values quantized in units of $1/p$.
Since $A_1^0$ vanishes, the covariant derivative in the $x_1$
direction is simply the partial derivative, and it follows that
\begin{equation}
k_1 = \frac{2 \pi m_1}{L_1} = \frac{2 \pi}{p} \frac{n_1}{L_1}; 
\qquad\qquad n_1 \in {\bb Z}.\label{k1}
\end{equation}

The boundary condition in the $x_2$ direction is trivial, indicating
that $m_2 \in {\bb Z}$. Due to the presence of the background,
however, the covariant derivative is modified by the commutator term.
One finds that
$$D_2\, \delta A(x_1,x_2) = \frac{1}{i} \partial_2 \delta A(x_1,x_2) -
[A^0,\delta A(x_1,x_2)] = \frac{2
\pi}{L_2}\left(m_2-\frac{r}{p}\right) \delta A (x_1,x_2).$$
Since $r$ could be any integer between 0 and $p-1$, we find that $k_2$
is also quantized in units of $1/p$:
\begin{equation}
k_2 = \frac{2 \pi}{p} \frac{n_2}{L_2}; \qquad\qquad n_2 \in {\bb Z} 
\label{k2}
\end{equation}

Thus, in general we find that the $p^2$ degrees of freedom in the
matrix appear as modes with momentum quantized in units of $1/p$ in
both the $x_1$ and $x_2$ directions.  Substituting the expressions for
$k_1$ and $k_2$ in equations (\ref{k1}) and (\ref{k2}) back into the
spectrum (\ref{back}), we recover
\begin{equation}
E^2 = \frac{1}{p^2+\left(\frac{4 \pi^2\alpha'}{L_1 L_2}\right)^2 q^2
}\left(\left(\frac{2\pi}{L_1}n_1\right)^2 + 
\left(\frac{2\pi}{L_2}n_2\right)^2 \right)
\end{equation}
in agreement with the string theory answer.  It is interesting to note
that the fractional quantization of momentum in the $x_1$ direction
appears naturally as a periodicity on a $p$-fold covering, whereas the
fractional quantization in the $x_2$ direction appears because there
are $p$ different ``types'' of excitations, labeled by which diagonal
of the matrix they appear in.  These two distinct mechanisms for
fractional quantization correspond naturally to the string geometry in
the D-brane picture, where the $x_1$ excitations correspond to
momentum on a $p$-fold wrapped brane while the $x_2$ excitations
correspond to strings connecting branes separated by a distance
proportional to $i-j$.

This explicit construction in the context of gauge theory gives an
interesting point of view on fractional momentum states, which have
been considered in a number of
contexts\cite{MaldaSuss96,DasMathur96,hashimoto96a}.
There is a close correspondence between this discussion and the
description of fractional momentum states as arising from the twisted
sector in an orbifold by the center of the gauge group
\cite{DMVV96,DVV97}.

\section{Intersecting Branes and Non-Abelian Born-Infeld}

In the previous section, we examined the D-brane/gauge theory
correspondence for tilted D-branes, corresponding to a gauge theory
background in the central $U(1)$ of $U(N)$. In this section, we
examine systems of intersecting D-branes which correspond to constant
curvature backgrounds with reducible connections.  When the constant
background field can be expressed in terms of a background connection
$A^0$ with components which all commute, the T-dual system corresponds
to a system of intersecting branes with well-defined positions.  For
concreteness, we will concentrate on systems corresponding to gauge
theories with $N=2$.  The constant curvature $U(2)$ backgrounds which
describe dual configurations of intersecting D-branes are precisely
those background fields which can be described in terms of reducible
connections.  Thus, in the gauge theory language, we will be
considering connections which decompose as a direct sum of $U(1)$
connections.  This restriction simplifies the analysis somewhat.  A
complete classification of constant curvature connections on $SU(2)$
bundles over $T^4$ was given by van Baal in \cite{vanBaal84}; in this
work he also calculated the Yang-Mills fluctuation spectra around
these backgrounds.  In the D-brane language, the corresponding
backgrounds describe a pair of branes intersecting at angles; the
spectrum of low-lying open strings in such backgrounds have been
studied in \cite{BDL96}.  Although the spectra computed in
\cite{vanBaal84} and \cite{BDL96} have striking qualitative
similarities, they are distinct in certain details. We will
investigate the possibility of resolving this discrepancy by
considering the full non-abelian Born-Infeld action.  We also discuss
the dynamics of non-supersymmetric backgrounds.

\subsection{Branes at angles\label{branes.at.angles}}

We begin our discussion by reviewing the D-brane construction of
\cite{BDL96}.  One of the main points made in \cite{BDL96} is the fact
that there exist supersymmetric configurations of branes intersecting
at angles.  Here, we consider a general class of configurations of
intersecting branes which includes non-supersymmetric brane
configurations.

The configurations we are interested in correspond to intersecting
D-branes living on tori.  As discussed above, these configurations are
dual to gauge theory backgrounds with reducible connections.  We will
here restrict attention to pairs of intersecting branes, concentrating
in particular on the case of two 2-branes on $T^4$.  There are several
kinds of string excitations around such a background.  There are
strings which stretch from each of the two 2-branes back to itself.
These strings carry momentum and winding numbers.  These excitations
correspond to modes living in one of the reducible components of the
dual bundle and are essentially equivalent to those discussed in the
previous section.  Thus, we will not consider these modes further here
but will concentrate on the strings connecting the two different
branes.  Unlike the strings connecting a brane to itself, the strings
connecting the different branes are unaffected by the compactification
of the space in which the branes are living.  This can be seen by
going to the covering space, where all strings connecting the two
branes are homotopically connected (as long as they are associated
with the same intersection point).  Of course, at higher order there
are diagrams connecting these strings with winding strings connecting
a brane to itself.  However, these diagrams will not affect the
fluctuation spectrum.  Thus, to understand the fluctuation spectrum of
the strings connecting the distinct 2-branes, it is sufficient to
understand the spectrum of strings connecting two 2-branes in
noncompact space.  This is precisely the situation considered in
\cite{BDL96}, and we recapitulate their discussion here with the
additional feature that we do not insist on the supersymmetry of the
D-brane configuration.

A sufficiently general configuration of two 2-branes in ${\bb R}^4$
can be defined by beginning with two 2-branes oriented along the 14
axis, and rotating one of them by an angle $\theta_1$ in the 12 plane
and by an angle $\theta_2$ in the 34 plane.  For $\theta_1 =
\theta_2$, this configuration was shown to be supersymmetric in
\cite{BDL96}. Following \cite{BDL96}, we use complex coordinates $Z_1$
and $Z_2$ instead of $X^1 + i X^2$ and $X^4 + i X^3$, respectively.

In world-sheet language, strings stretching between the two branes are 
described by open strings with boundary conditions
\begin{eqnarray*}
\Re {\partial\over\partial\sigma} Z^i|_{\sigma=0} &=& 0 \\
\Im Z^i|_{\sigma=0} &=& 0 \\
\Re e^{i\theta_i}{\partial\over\partial\sigma} Z^i|_{\sigma=\pi}& =& 0 \\
\Im e^{i\theta_i}Z^i|_{\sigma=\pi} &=& 0 .
\end{eqnarray*}
The mode expansion for the complex bosonic field  is given by
$$Z^i(w,\bar w)=\sum_{m\in \BZ} \left\{ x^i_{-\alpha_i +m}e^{i(m-\alpha_i )w}
+\tilde x^i_{\alpha_i+m}e^{-i(m+\alpha_i)\bar w}\right\}$$
where $w
= \sigma +  \tau, \bar{w} = \sigma - \tau$ and $\alpha_i =
\theta_i/\pi$.  The vacuum energy
$$\Delta E = -\frac{1}{2} + \frac{\alpha_{1}}{2}
+ \frac{\alpha_{2}}{2}$$
can be computed either from the $\zeta$-function regularization of
the sum
\begin{eqnarray*}
\alpha + (\alpha+1) + (\alpha+2) + \ldots &=& -\frac{1}{12}
 + \frac{1}{8} \alpha\\
(\frac{1}{2}+\alpha) + (\frac{3}{2}+\alpha) + \ldots & =& \frac{1}{24}
 - \frac{1}{8} \alpha 
\end{eqnarray*}
or from the dimensions of the world sheet operators needed to twist
the boundary condition \cite{BDL96,hashimoto96,orbifold}.  The
low-lying open string excitations in the NS sector described in
equation (3.5) of \cite{BDL96} generalize to
$$
\begin{array}{rr}
\tilde\psi^i_{\alpha_1-1/2} (x_{-\alpha_1}^1 )^{n_1}
(x_{-\alpha_2}^2 )^{n_2} | 0 \rangle; &
 \tilde\psi^i_{\alpha_2-1/2} (x_{-\alpha_1}^1 )^{n_1}
(x_{-\alpha_2}^2 )^{n_2} | 0 \rangle\\
\psi^i_{-\alpha_1-1/2} (x_{-\alpha_1}^1 )^{n_1}
(x_{-\alpha_2}^2 )^{n_2} | 0 \rangle; &
\psi^i_{-\alpha_1-1/2} (x_{-\alpha_1}^1 )^{n_1}
(x_{-\alpha_2}^2 )^{n_2} | 0 \rangle\\
&\psi^\mu_{-1/2} (x_{-\alpha_1}^1 )^{n_1}
(x_{-\alpha_2}^2 )^{n_2} | 0 \rangle\\
\end{array}
$$
and their contribution to the masses are, respectively,
\begin{eqnarray}
(-\alpha_1+1/2) + n_1 \alpha_1 + n_2 \alpha_2 + \Delta E & = &
(n_1-1/2)\alpha_1 + (n_2+1/2) \alpha_2 \nonumber \\ 
(-\alpha_2+1/2) + n_1 \alpha_1 + n_2 \alpha_2 + \Delta E & = &
(n_1+1/2)\alpha_1 + (n_2-1/2) \alpha_2 \nonumber \\ 
(\alpha_1+1/2) + n_1 \alpha_1 + n_2 \alpha_2 + \Delta E & = & 
(n_1+3/2)\alpha_1 + (n_2+1/2) \alpha_2 \label{eq3} \\
(\alpha_2+1/2) + n_1 \alpha_1 + n_2 \alpha_2 + \Delta E & = &
(n_1+1/2)\alpha_1 + (n_2+3/2) \alpha_2 \nonumber \\ 
1/2 + n_1 \alpha_1 + n_2 \alpha_2 + \Delta E & = & (n_1+1/2)\alpha_1 +
(n_2+1/2)\alpha_2  \nonumber 
\end{eqnarray}
This gives a complete characterization of the fluctuation spectra of
strings stretching between two 2-branes at angles $\theta_1,
\theta_2$, including cases where the brane configuration is not BPS.
Note that when $\alpha_1 \neq \alpha_2$ the spectrum contains a
tachyon.

\subsection{Yang-Mills fluctuation spectra}

We will now discuss a gauge theory configuration whose dual
corresponds to two 2-branes on $T^4$.  As stated, we will restrict
attention to $U(2)$ bundles with connections which are reducible to a
direct sum of two $U(1)$ connections.  The discussion could be
generalized in a straightforward way to arbitrary reducible $U(N)$
bundles.  It was pointed out by 't Hooft \cite{tHooft81} that for $N >
2$ there are constant curvature connections which do not satisfy
abelian boundary conditions; these correspond to D-brane
configurations with fewer than $N$ D-branes, with some of the D-branes
wrapped multiple times around the base space as in Section \ref{sec:pq}.

\begin{figure}
\centerline{\psfig{file=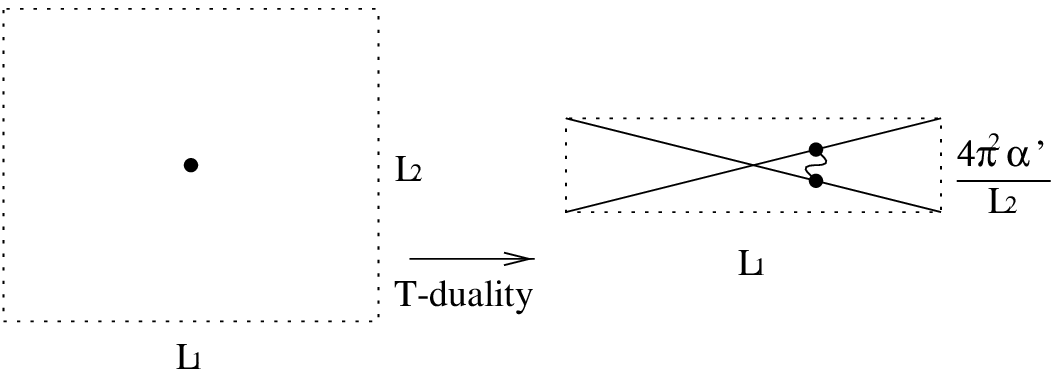}}
\newcaption{(12)-plane cross section of two 4-branes with a constant field
strength 2-instanton background and the T-dual intersecting 2-branes.
\label{inter1}}
\end{figure}

We will consider here a constant background field given by a $U(2)$
connection $A^0_\mu$ on $T^4$ with the following components
\begin{eqnarray*}
 A^0_1 &=& 0 \\
 A^0_2 &=& \frac{\pi}{L_2 L_1} n_{21} x_1 \tau_3\\
 A^0_3 &=& \frac{\pi}{L_3 L_4} n_{34} x_4 \tau_3\\
 A^0_4 &=& 0.
\end{eqnarray*}
where $L_\mu$ are the dimensions of the 4-torus and $n_{\mu \nu}$ is
an anti-selfdual matrix with with $n_{21} = -n_{12}= n_{34} = -n_{43}=
2$ and zero otherwise.  $\tau_3$ is the SU(2) generator $\rm
Diag\{1,-1\}$.  This configuration has an instanton number
\cite{vanBaal82}.
$$C_2 = \frac{n_{\mu \nu} \tilde n_{\mu \nu}}{8}=2$$
Because the field strength is proportional to $\tau_3$ the first Chern
class vanishes.  Thus, this configuration corresponds to a (singular)
point in the moduli space of $U(2)$ instantons on $T^4$ with instanton
number $k = 2$.  (Note that while a single instanton on $T^4$ shrinks
to a point, there is a regular moduli space of instantons with $k = 2$.)
Performing a T-duality transformation along directions $2$ and 3 will
invert the periods along these directions, and will give rise to a
pair of 2-branes embedded according to the equations (see figure
\ref{inter1})
\begin{eqnarray*}
X_2 &=& 2 \pi \alpha'  A^0_2 = \pm  \frac{2 \pi^2 \alpha'}{L_1
L_2} n_{21}x_1 \\ 
X_3 &=& 2 \pi \alpha'  A^0_3 = \pm  \frac{2 \pi^2 \alpha'}{L_3
L_4} n_{34}x_4  
\end{eqnarray*}
The field strength $F_{\mu \nu}$ can immediately be related to the
angles $\theta_1, \theta_2$ between the branes:
\begin{eqnarray*}
\tan(\theta_1/2) &=& 2 \pi \alpha' F_{21}\\
\tan(\theta_2/2) &=& 2 \pi \alpha' F_{34}
\end{eqnarray*}
The connection $A^0_\mu$ is gauge equivalent to one considered in
\cite{vanBaal84},
$$A_\mu(x) = -\frac{1}{2} \frac{\pi }{L_\mu L_\nu} n_{\mu \nu} x_\nu \tau_3.$$
Van Baal investigated the spectrum of $SU(2)$ Yang-Mills
fluctuations around backgrounds of this type.  
He explicitly expanded the Yang-Mills action around $A^0$ using
$$A_\mu = A_\mu^0 + \left(
\begin{array}{cc}
b_\mu^1 & \sqrt{2} c_\mu^* \\
\sqrt{2} c_\mu  & b_\mu^2
\end{array}
\right)
$$
to second order in $b_\mu$ and $c_\mu$. Here, $b_\mu$ and $c_\mu$ are
real and complex fields respectively; in $SU(2)$, we would have
$b_\mu^1 = -b^2_\mu$.  There are also ghost fields $\psi$ living in
the adjoint representation.  The action, expressed in these variables,
takes the form
$$S = S_0 + \int d^4 x \frac{1}{2} b_\mu^a (M_0 \delta^{\mu \nu}
\delta^{ab}) b_\nu^b + c_\mu^* (M_n \delta^{\mu \nu} - 4 \pi i F^{\mu
\nu}) c_\nu + {\rm Tr}\; (\psi^{\dagger} M_{gh} \psi) + {\cal
O}(\delta A_\mu^3)$$
with
\begin{eqnarray*}
M_0 &=& \left( \frac{1}{i} \partial_\mu \right)^2 \\
M_n &=& \left( \frac{1}{i} \partial_\mu - \pi F_{\mu\nu} x_\nu \right)^2\\
M_{gh} & = & -D_\lambda^2
\end{eqnarray*}
As discussed in Section \ref{branes.at.angles}, the fields $b^i$
correspond to strings in the dual picture connecting a brane with
itself, and have spectra corresponding to the discussion in Section
\ref{sec:tilted}.  Thus, we concentrate now on the spectrum for the
field $c$ corresponding to strings stretching between the two branes.
The computation of this spectrum is described in detail in
\cite{vanBaal84}, so we simply summarize the results of that
investigation here\footnote{Although van Baal was studying the purely
bosonic four-dimensional theory, and thus did not include the adjoint
scalars associated with transverse fluctuations of the D-brane
world-volume in his analysis, the quadratic forms in the ghost kinetic
terms are identical in structure to the quadratic forms of these
adjoint scalars and can be used to compare their spectra.}.  The
boundary conditions on $c$ indicate that it is a theta function on
$T^4$.  The operator $M_n$ acts on the appropriate space of theta
functions as a Laplace-Beltrami operator up to a constant.  The
eigenfunctions of this operator were classified in \cite{Hano}, and
are given by a sum over the lattice of harmonic oscillator
eigenfunctions rotated by linear phases.  The low-lying eigenfunctions
are localized in the vicinity of the string intersection points.  We
will discuss a specific example of such eigenfunctions more explicitly
in Section \ref{sec:unstable}.  The spectrum of eigenvalues for the
four-dimensional system of interest here is given in equation (3.25) of
\cite{vanBaal84}:
$$
\begin{array}{rr}
M_n -4 \pi iF
   :   &  2 \pi  (2m_1 -1)  f_1  + 2 \pi(2m_2+1)  f_2 \\
       &  2 \pi  (2m_1 +1)  f_1  + 2 \pi(2m_2-1)  f_2 \\
       &  2 \pi  (2m_1 +3)  f_1  + 2 \pi(2m_2+1)  f_2 \\
       &  2 \pi  (2m_1 +1)  f_1  + 2 \pi(2m_2+3)  f_2 \\
M_{gh}:&  2 \pi  (2m_1 +1)  f_1  + 2 \pi(2m_2+1)  f_2 \\
\end{array}
$$
where 
\begin{eqnarray*}
f_1 &=& \frac{1}{ \pi} F_{21} = \frac{\tan(\theta_1/2)}{2 \pi^2 \alpha'}\\
f_2 &=& \frac{1}{ \pi} F_{34} = \frac{\tan(\theta_2/2)}{2 \pi^2 \alpha'}
\end{eqnarray*}
We therefore see that the spectrum of fluctuations around this constant
field strength background has
precisely the structure of the oscillator excitations (\ref{eq3}) found from
the string theory point of view in \cite{BDL96} up to the
identification
\begin{eqnarray*}
\frac{2\tan(\theta_1/2)}{\pi} &\leftrightarrow& \alpha_1 =
\frac{\theta_1}{\pi}\\ 
\frac{2\tan(\theta_2/2)}{\pi} & \leftrightarrow& \alpha_2 =
\frac{\theta_2}{\pi} 
\end{eqnarray*}
\begin{figure}
\centerline{\psfig{file=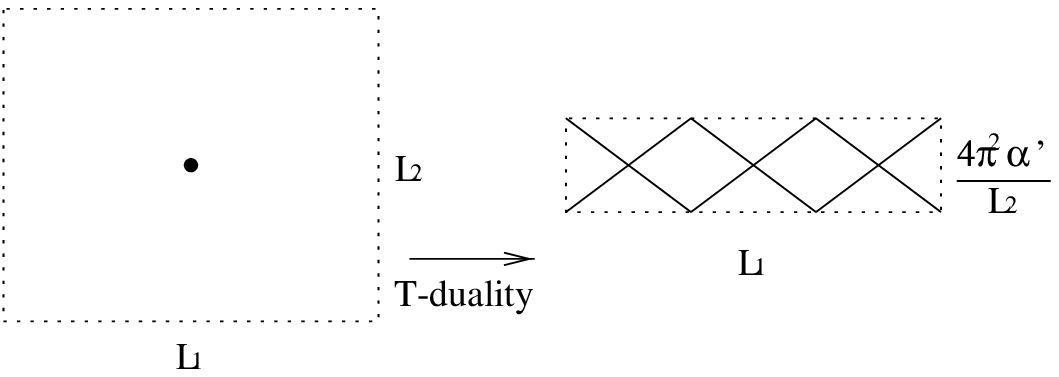}}
\newcaption{(12)-plane cross section of two 4-branes with a constant
field strength  many-instanton background and the T-dual intersecting
2-branes with multiple winding.
\label{inter2}}
\end{figure}
Thus, the fluctuation spectra computed from these two points of view
agree up to a discrepancy between $\theta/2$ and $\tan
(\theta/2)$.  For small $\theta$ these spectra agree exactly.


\subsection{Resolving the discrepancy using non-abelian Born-Infeld}

The difference in the spectra resulting from the discrepancy between
$\theta/2$ and $\tan (\theta/2)$ might seem minor in light of the fact
that they are in perfect agreement at small angles, and that in the
field theory limit keeping the background constant, $\theta$ goes to
zero.  However, just as in Section \ref{sec:tilted}, one can take a
field theory limit while scaling the background to keep $\theta$
finite but small compared to 1.  To be more specific, we can consider
the background
$$A_\mu(x) = -\frac{1}{2} \frac{\pi }{L_\mu L_\nu} q n_{\mu \nu}
x_\nu \tau_3,\qquad q \in \BZ$$ 
and take the limit $\alpha'/L_1^2 \rightarrow 0$ keeping the ratios of
$L_\mu$'s and $\sqrt{q \alpha'}$ constant. After a gauge
transformation to set $A_1 = A_4 = 0$, this will lead in the T-dual
picture to 2-branes wrapping multiply around the 23 cycles as
illustrated in figure \ref{inter2}. 

We saw in Section \ref{sec:tilted} that scaling the field strength
enhances the terms ordinarily subleading in $\alpha'$ and that the
full Born-Infeld action was needed to recover the exact correspondence
in the case of tilted branes.  It is therefore natural to suspect that
the same mechanism underlies the resolution of the $\theta/2$ vs.\
$\tan(\theta/2)$ discrepancy.  We investigate this possibility in this
subsection.  The analysis is complicated by the fact that the
background field does not commute with all fluctuations.  The field
theory analysis therefore depends on the full structure of the
non-abelian Born-Infeld action.  Recently, a concrete formulation of
the non-abelian Born-Infeld action was proposed by Tseytlin
\cite{NDBI}.  Using Tseytlin's Born-Infeld action, we compute the
spectrum of fluctuations around the same background as was considered
in the previous subsection.  A remarkable relation emerges in the
analysis, bringing the field theory spectrum tantalizingly close to
the string theory spectrum of section \ref{branes.at.angles}.
However, we fall slightly short of reproducing the string theory
spectrum in its entirety.  We interpret this situation in the
following manner: Due to the highly non-trivial nature of the partial
success in resolving the $\theta/2$ vs.\ $\tan (\theta/2)$
discrepancy, there must be a strong element of truth in Tseytlin's
formulation of the non-abelian Born-Infeld action, but in as much as
it fails to reproduce the string theory spectrum precisely, it is
still in some sense incomplete.  Reformulating the Born-Infeld action
to resolve this discrepancy is likely to be a rather tricky
enterprise.  Nonetheless, we feel that reproducing the spectrum of a
system of D-branes intersecting at an angle will serve as a useful
test and could possibly be used as a guiding principle in formulating
the elusive non-abelian Born-Infeld action.

The key feature of Tseytlin's non-abelian Born-Infeld action is
its resolution of product ordering ambiguities through symmetrization
$$\lag = {\rm STr} \sqrt{-\det(\eta_{\mu \nu} + 2 \pi \alpha' F_{\mu \nu})}.$$
Our strategy here is to perform formal manipulations of quantities
inside the symmetric trace as if they were abelian. At the last stage
of the analysis, we will re-insert the non-commuting factors and
perform the symmetrized trace, thereby promoting the abelianized
action back up to its original non-abelian counterpart. 
In our analysis we consider the symmetrized trace term by term in an
expansion with respect to the background as well as the fluctuation. For
example, a term in the expansion corresponding to an abelianized expression,
$$(F_0)^m  X^2,$$
where $F_0 = F_0^3 \tau_3$ and $X = \sum X^a \tau_a$ are in the
adjoint of $SU(2)$,
is replaced by
\begin{equation}
{\rm STr} (F_0 \tau_3) ^m (\tau_a X^a) (\tau_b X^b) =
F_0^m X^a X^b \left[\frac{1}{m+1} \sum_{i=0}^m{\rm Tr}
(\tau_3^i \tau_a \tau_3^{m-i} \tau_b)\right]
\label{expansion}
\end{equation}
upon reinsertion of non-commuting factors.  Let us consider in
particular the case where $a$ and $b$ are restricted to the range of
values $\{1, 2\}$.  Simple algebraic manipulations of the Pauli
matrices show that
$${\rm Tr}
(\tau_3^i \tau_a \tau_3^{m-i} \tau_b) = 
{\rm Tr}
((-1)^i  \tau_3^m \tau_b \tau_a) $$
At this point, it is convenient to consider the cases of $m$ even and
$m$ odd separately.
For $m$ odd,
$$\sum_{i=0}^m{\rm Tr}((-1)^i  \tau_3^m \tau_b \tau_a) =
\sum_{i=0}^m{\rm Tr}((-1)^i  \tau_3 \tau_b \tau_a) = 0$$
whereas for $m$ even,
$$\sum_{i=0}^m{\rm Tr}((-1)^i  \tau_3^m \tau_b \tau_a) =
\sum_{i=0}^m{\rm Tr}((-1)^i \tau_b \tau_a) = \delta_{ab} $$
Therefore, if in the abelian Born-Infeld action the terms quadratic in
fluctuations had an expansion consisting solely of terms with even powers
in the background: 
$$S_{BI}(F_0, \tilde{F}) = (\tilde{F}^2) \left( \rule{0ex}{3ex} a_0 +
a_1 F_0^2 + a_2 F_0^4 + \ldots a_m F_0^{2m}+ \ldots, \right)$$ 
then its non-abelian generalization for $\tau_1$ and $\tau_2$ fields
will give rise to an expansion of the form
$$S_{NBI}(F_0, \tilde{F}) = \delta_{ab} (\tilde{F}^a \tilde{F}^b) \left(
\rule{0ex}{3ex} a_0 + \frac{a_1}{3} F_0^2 + \frac{a_2}{5} F_0^4 +
\ldots+  \frac{a_m}{2m+1} F_0^{2m}+ \ldots \right).$$ 
These two expansions are related by
\begin{equation}
S_{NBI}(F_0,\tilde{F}) = \frac{1}{F_0} \int_0^{F_0} d \bar F_0
S_{BI}(\bar F_0, \tilde{F}) 
\label{relation}
\end{equation}

The goal at this point is to examine the fluctuations of
$$
\lag = {\rm STr} \sqrt{-\det(\eta_{\mu \nu} + 2 \pi\alpha' F^0_{\mu
\nu}+2 \pi \alpha' \tilde{F}_{\mu \nu})} 
$$
in a  background defined by
$$A_\mu = A^0_\mu +   \delta A_\mu$$
and
$$
\tilde{F}_{\mu \nu}  =  
\partial_\mu   \delta A_\nu - \partial_\nu \delta A_\mu
 -i [ \delta A_\mu, A^0_\nu] -i[ A^0_\mu,\delta A_\nu]
-i[ \delta A_\mu,   \delta A_\nu]
$$
We expand the action to quadratic order using the standard identity
$$
\sqrt{\det( M_0 + \delta M)} = \sqrt{\det (M_0)}
\left( 1 + \frac{1}{2} \tr[M_0^{-1} \delta M] + \frac{1}{8}
(\tr[M_0^{-1} \delta M])^2 - \frac{1}{4} \tr[M_0^{-1} \delta M M_0^{-1}
\delta M] \right) 
$$
In terms of our variables, the quadratic action reads
\begin{eqnarray}
\lag_2 &=& {\rm Str} \left[\sqrt{-\det( \eta_{\mu \nu} + 2 \pi \alpha' F^0_{\mu
\nu})} 
\left\{\frac{1}{2} B^{\mu \nu} \tilde F_{\nu \mu} - \frac{1}{4}
g^{\mu \nu} \tilde{F}_{\nu \lambda} g^{\lambda \sigma}\tilde{F}_{\sigma \mu}  
\right.\right.\nonumber \\
&&\left.\left.\qquad\qquad\qquad\qquad\qquad
-\frac{1}{4}\left( B^{\mu \nu} \tilde{F}_{\nu \lambda} B^{\lambda \sigma}
\tilde{F}_{\sigma \mu} - \frac{1}{2} \left(B^{\mu \nu} \tilde{F}_{\mu
\nu} \right)^2 \right) \right\}  \right]
\label{quadaction}
\end{eqnarray}
where now,
$$
g^{\mu \nu} = (\eta_{\mu \nu} +2 \pi \alpha' F^0_{\mu \nu})^{-1}_{\rm sym} = {\rm Diag}
\{-1, g_{(12)} , g_{(12)}, g_{(34)}, g_{(34)} \}
$$
with
\begin{eqnarray*}
g_{(12)} & = & \frac{1}{1+(2 \pi \alpha' F_{21})^2}\\
g_{(34)} & = & \frac{1}{1+(2 \pi \alpha' F_{34})^2}
\end{eqnarray*}
and 
$$B^{\mu \nu} = 
-g^{\mu \lambda} F_{\lambda \rho} \eta^{\rho \nu}.$$
The term linear in $\tilde F$ contributes at quadratic order because
the commutator term in $\tilde F$ is quadratic in the fluctuations.
This time, the terms quadratic in $B^{\mu \nu}$ turn out not to vanish
but give a topological contribution
$$
B^{\mu \nu} \tilde{F}_{\nu \lambda} B^{\lambda \sigma} \tilde{F}_{\sigma \mu}
- \frac{1}{2} \left(B^{\mu \nu} \tilde{F}_{\mu \nu} \right)^2 = \frac{2F_{21}
F_{34}}{(1+(2 \pi \alpha'F_{21})^2)(1+(2 \pi \alpha' F_{34})^2)} \epsilon^{\mu \nu \lambda \sigma}
\tilde{F}_{\mu \nu} \tilde{F}_{\lambda \sigma}
$$
which can be ignored in the discussion of the fluctuation spectrum.

In order to compute the fluctuation spectrum we now must apply the
integral prescription (\ref{relation}) to the action in order to deal
with the lack of commutativity between the fluctuations and the
background.  Tseytlin's symmetrized trace prescription indicates that
we should associate a single $U(2)$ generator with each factor of the
field strength $F$.  This gives a well-defined formulation of a
non-abelian action.  With this prescription, the action as a function
of the components of the field strength and its off-diagonal
fluctuations should be
computed by applying (20) to the terms quadratic in $\tilde{F}$ in
(21).  The term linear in $\tilde{F}$ is not modified, since it is
only nonzero when $\tilde{F}$ commutes with the background.
Unfortunately, this does not seem to give exactly the
results that we would expect from string theory.  Nonetheless, there
is some indication that this prescription comes close to doing the
right thing.  To see this let us consider a special case where the
field strength is anti-self-dual, corresponding to a BPS brane
configuration.  In this case we have $F_{21} = F_{34}$.  The
determinant factor in the action gives a single factor of
$g_{(12)}^{-1}$.  This cancels with the $g$ factors in the terms
quadratic in $\tilde{F}_{0i}$, so that the kinetic terms are
canonically normalized.  The potential terms quadratic in $\tilde{F}$
carry two factors of $g$ so that they are multiplied by an overall
factor of $g_{(12)}$ compared to the pure Yang-Mills theory.

At this point, we encounter a pleasant surprise. Applying the relation
(\ref{relation}) to the metric leads to
%
\[
\frac{1}{F_{21}} 
\int_0^{F_{21}} d \bar F\, g_{(12)}   =
\frac{1}{F_{21}} 
\int_0^{F_{21}} d \bar F \frac{1}{1+(2 \pi \alpha' \bar F)^2}  =  
\frac{\tan^{-1}(2 \pi \alpha' F_{21})}{F_{21}} 
= \frac{\theta_1/2}{\tan(\theta_1/2)} 
\]
This is precisely the rescaling necessary to match the field theory
spectrum with the string theory result!  In fact, in this case we find
that the spacing of the fluctuation spectrum corresponds precisely to
that predicted by string theory.  Unfortunately, however, the energy
of the lowest fluctuation is also affected by the term linear in
$\tilde{F}$.  The factors of $g$ cancel in this term so that the
calculation of standard Yang-Mills theory is not modified by the
symmetrized trace form of the Born-Infeld action.  Thus, even in the
special case where the field is anti-self-dual, we do not get exactly
the spectrum predicted by string theory.

We would expect that even in the more general case where the field is
not anti-self-dual, the spectra computed from string theory and from the
non-abelian Born-Infeld theory should agree.  In the more general
case, however, we find that even the spacing of the energy levels
fails to be reproduced exactly by the symmetrized trace prescription.
Thus, it seems clear that this specific formulation of the non-abelian
Born-Infeld theory is not sufficiently general to capture all the
relevant physics of the situation we are considering here.

There are several possible explanations for why the symmetrized trace
prescription does not exactly match the string theory calculation.
One possibility is that the association of a single  generator
to each factor of the field strength should be modified.  As an
example of this sort of modification, we might choose to normalize the
fluctuations $\delta A$ so that the kinetic terms in the action always
have a canonical normalization.  For cases where the field strength is
not anti-self-dual this will incorporate extra factors into the
potential terms.  It can be shown that these extra terms guarantee
that the scaling factor associated with the quadratic term in
$\tilde{F}$ will give fluctuations with a level spacing precisely that
predicted by string theory, as we showed above explicitly in the
anti-self-dual case.  Unfortunately, however, it does not seem to be
possible to arrange the normalization so that the linear term in
$\tilde{F}$ also scales correctly.

Another possible reason why the symmetrized trace prescription does
not reproduce the string theory results correctly is that covariant
derivative terms $DF$ or commutator terms $[F, F]$ may be relevant in
this situation.  In Tseytlin's discussion of the non-abelian
Born-Infeld action, terms of these forms were explicitly dropped.
Since the fluctuations we are concerned with here are theta functions
which give nontrivial spatially dependent fluxes, it is quite possible
that the effects of covariant derivative terms must be incorporated
into the non-abelian Born-Infeld action in order to correctly
reproduce the string theory results.  A discussion of a related
discrepancy was given in \cite{Argyres-Nappi}, where it was shown that
it was necessary to add commutator and covariant derivative terms to
the Born-Infeld action already at order $F^4$ to reproduce results
from string theory.

In any case, we do not have a complete understanding of why the
prescription we have described corrects some, but not all, of the
terms appearing in the action, and as such we have not provided a
satisfactory resolution to the problem of the $\theta/2$ vs.\ $\tan
(\theta/2)$ discrepancy. Nonetheless, the relation between $\theta/2$
and $\tan (\theta/2)$ is transcendental, and subtle details must work
out just right in order for the correct transformation to emerge from
this kind of analysis.  The fact that one can naturally derive from
the symmetrized trace prescription the precise scaling factor needed
to achieve agreement with string theory seems to provide some evidence
that there is an element of truth in this description.

We believe that when it is correctly defined, the full non-abelian
Born-Infeld action should resolve the discrepancy we have discussed
here.  We therefore pose the resolution of the $\theta/2$ vs.\ $\tan
(\theta/2)$ discrepancy as a test which should be satisfied by any
candidate for the full non-abelian Born-Infeld action.

\subsection{Brane-anti-brane configurations}
\label{sec:unstable}

We have discussed the correspondence between fluctuation spectra in a
variety of situations corresponding to tilted and intersecting
D-branes.  As mentioned in Section \ref{branes.at.angles}, the
formulae for the fluctuation spectra around intersecting D-branes
extend to cases where the configuration of branes breaks
supersymmetry.  However, in such situations the state is no longer BPS
and generally admits a tachyonic instability \cite{bankssusskind}.  In
this subsection we discuss the simplest example of such an instability
in slightly more explicit terms.

\begin{figure}
\psfig{file=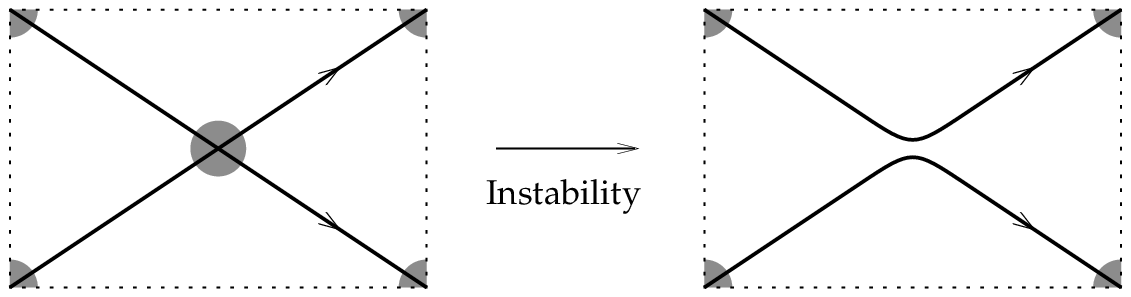} \newcaption{Two intersecting 1-branes on
$T^2$ and  their geometric instability.  The regions around the two
intersection points where the tachyonic modes are localized are shaded.
\label{f:unstable}}
\end{figure}

Consider a pair of 1-branes wrapped on $T^2$ with winding numbers
$(1,1)$ and $(1, -1)$.  These 1-branes intersect at two points.  This
configuration is unstable, since a simple change in the topology at
one of the crossing points turns the configuration into a single
1-brane with winding $(2, 0)$ (see Figure~\ref{f:unstable}).  We can
understand this instability in the system from both the string theory
and the Yang-Mills/Born-Infeld point of view.  In the gauge theory
picture, we have a pair of 2-branes on $T^2$, one carrying a unit of
0-brane charge, and the other carrying a negative 0-brane charge,
corresponding to a single anti-0-brane.  Thus, the tachyonic instability in
this picture corresponds to the instability of a brane-anti-brane
configuration.

To be specific, we will consider a $U(2)$ gauge theory on a $T^2$ with
sides of length $L_1, L_2$.  We will take the boundary conditions to
be given by
\begin{eqnarray*}
\Omega_1 (x_2) & = &  e^{2 \pi i (x_2/ L_2) \tau_3}\\
\Omega_2 (x_1) & = &  I
\end{eqnarray*}
Although these seem like nontrivial boundary conditions, they define
a trivial $U(2)$ bundle over $T^2$ in an unusual choice of gauge; this
gauge choice allows us to write the background connection in a form
which manifestly corresponds to the desired T-dual D-brane configuration.
The boundary conditions on the gauge fields are given by
\begin{eqnarray*}
A_1 (x_1 + L_1, x_2) & = &  e^{2 \pi i (x_2/L_2) \tau_3} 
 A_1 (x_1, x_2)  e^{-2 \pi i (x_2/L_2) \tau_3}\\
A_1 (x_1, x_2 + L_2) & = &    A_1 (x_1, x_2) \\
A_2 (x_1 + L_1, x_2) & = &  e^{2 \pi i (x_2/L_2) \tau_3}  A_2 (x_1, x_2) 
 e^{-2 \pi i (x_2/L_2) \tau_3}
  + \left(\frac{2 \pi}{L_2 }\right) \tau_3  \\
A_2 (x_1, x_2 + L_2) & = &    A_2 (x_1, x_2)
\end{eqnarray*}
We will be considering the
constant curvature background corresponding to the connection
\begin{eqnarray*}
A^0_1 & = &  0\\
A^0_2 & = &  \frac{2 \pi}{L_1 L_2}  x_1 \tau_3.
\end{eqnarray*}
Clearly, through T-duality this corresponds to a pair of 1-branes
whose transverse positions are given by
\[
X_2 =  \pm \frac{4 \pi^2 \alpha'}{L_1 L_2}  x_1,
\]
just as shown in figure~\ref{f:unstable}.

The tachyonic instability of this system is easy to see in the D-brane
language.  The string excitation spectrum is essentially given by
(\ref{eq3}) where we set $\alpha_2 = n_2 = 0$ and where $\alpha_1$ is
related to the angle $\theta$ between the 1-branes by $\alpha_1 =
\theta/\pi$.  As mentioned in section \ref{branes.at.angles}, this
spectrum contains a tachyon.

We now discuss the instability of this system from the gauge theory
point of view.  As we have seen in previous sections, the Born-Infeld
action is necessary to get the precise normalization of the
fluctuation spectra.  However, the sign of the various fluctuations
are not dependent upon such a precise calculation, so we should be
able to detect the instability within the simple Yang-Mills framework.
To find the unstable modes, we can use essentially the same analysis
as that of van Baal \cite{vanBaal84}, simply setting to zero one of
the two curvatures $f_1, f_2$.  We find that indeed there are modes
with negative eigenvalues corresponding to unstable directions in the
system.  To describe these modes more precisely, let us consider the
boundary conditions on the off-diagonal modes of the system.
Decomposing as above
$$A_\mu = A_\mu^0 + \left(
\begin{array}{cc}
b_\mu^1 & \sqrt{2} c_\mu^* \\
\sqrt{2} c_\mu  & b_\mu^2
\end{array}
\right)
$$
we find that each $c_\mu$ satisfies the boundary conditions
\begin{eqnarray}
c_\mu (x_1 + L_1, x_2) & = &  e^{- 4 \pi i (x_2/L_2)} c_\mu (x_1, x_2)
\nonumber\\
c_\mu (x_1, x_2 + L_2) & = &  c_\mu (x_1, x_2)\label{eq:boundary}.
\end{eqnarray}
These conditions imply that $c_{\mu}$ is a section of a $U(1)$ bundle
with $C_1 = 2$.  
We can express $c_\mu$  as a theta function satisfying the
condition
\[
\Theta (z + q) = e^{2 \pi q \bar{z} + \pi | q |^2} \Theta (z)
\]
where $q \in \BZ+ i\BZ$
by transforming
\[
c_{\mu} (x_1, x_2) = \exp \left(\frac{2 \pi i x_1 x_2}{L_1 L_2}
  -\pi (\frac{x_1^2}{L_1^2} + \frac{x_2^2}{L_2^2} )\right) 
\Theta (x_1/L_1 + ix_2/L_2).
\]
A general discussion of the theta
functions of this type which are eigenfunctions of the
Laplace-Beltrami operator is given in \cite{Hano}.  In the particular
case of interest here, the description of the states in terms of the boundary
conditions (\ref{eq:boundary}) is fairly straightforward.
We are looking for  eigenfunctions of the operator
\[
M_n = -\partial_1^2 +(i \partial_2
 -  \frac{4 \pi x_1}{L_1 L_2})^2 - \frac{8\pi}{L_1 L_2}
\]
with the minimal eigenvalue, subject to the condition that the
eigenfunctions must satisfy the boundary conditions
(\ref{eq:boundary}).  There are precisely two such eigenfunctions with
the minimum eigenvalue $-4 \pi/(L_1 L_2)$.  These functions are given by
\[
c(x_1, x_2) = 
\sum_{n, m \in\BZ}  \exp \left(
-  \frac{\pi}{L_1 L_2} [ (x_1-L_1 n)^2  + (x_2-L_2 m)^2]
+ 2 \pi 
i (\frac{x_1}{L_1} m-\frac{x_2}{L_2} n)
-2 \pi i \frac{x_1 x_2}{L_1 L_2} 
\right).
\]
and
\begin{eqnarray*}
c'(x_1, x_2)  & = & 
\sum_{\hat{n}, \hat{m} \in\BZ+ 1/2}  \exp \left(
 \pi i (\hat{n} + \hat{m})
-  \frac{\pi}{L_1 L_2} [ (x_1-L_1 \hat{n})^2  + (x_2-L_2 \hat{m})^2]
\right.\\
& &\hspace{1.3in}\left.+ 2 \pi 
i (\frac{x_1}{L_1} \hat{m}-\frac{x_2}{L_2} \hat{n})
-2 \pi i \frac{x_1 x_2}{L_1 L_2} 
\right).
\end{eqnarray*}

These solutions are given by superimposing copies of the gaussian
harmonic oscillator ground state centered around points on a lattice.
As $L_2$ becomes small, these functions become  localized around
the values $x_1 = 0$ and $x_1 =L_1/2$, which are precisely the $x_1$
values of the two points of intersection for the dual D-branes.  In
the corresponding D-brane picture, as $L_2$ becomes small the dual
radius becomes large and the angle between the branes increases, which
as discussed in \cite{BDL96} should localize the string modes in the
vicinity of the intersection points.  Furthermore, since the unstable
modes are associated with the $c$ entries of the $U(2)$ matrix, which
affect the commutativity between the two brane positions, it is
natural to interpret these modes geometrically in terms of a change in
topology localized to the brane intersection points.  This corresponds
perfectly with what we would expect from the geometrical picture shown
in figure~\ref{f:unstable}.  How this explicit description of the
unstable mode relates to the D-brane instability is difficult to make
completely precise, however, due to the nonlocality of the T-duality
transformation in the $x_2$ direction.  A better understanding of this
relationship might lead to interesting results about brane-anti-brane
interactions which could be understood from a simple field theory
perspective.

\section{Conclusions}

In this paper, we examined the spectra of low-energy fluctuations
around various D-brane configurations in string theory and the T-dual
spectra of fluctuations around constant background fields in gauge
theory.  

Background fields in the central $U(1)$ of a $U(N)$ gauge theory
correspond to tilted branes on the dual torus.  By taking the field
theory limit while keeping the tilt angle constant, we encountered a
discrepancy between the field theory spectrum and the string theory
spectrum.  The discrepancy was resolved by computing the fluctuation
spectrum of the full Born-Infeld action instead of the Yang-Mills 
action. This is a natural result in light of the fact that the scaling
limit which keeps the tilt constant corresponds to scaling up the background
field strength keeping $2 \pi \alpha' F$ constant. This forces the
terms ordinarily subleading in $\alpha'$ to play a relevant role.

We also examined the spectrum of excitations around constant
background fields from reducible connections, corresponding to branes
intersecting at an angle on the dual torus.  We found that the results
of \cite{BDL96} can be extended to non-supersymmetric configurations
and that the dynamics of tachyons can be studied naturally in the
context of gauge theories.  In the small angle limit, the D-brane and
Yang-Mills spectra are in agreement.  The correspondence breaks down,
however, when the angle of intersection is fixed and finite.  We
investigated the possibility of making the correspondence exact even
at finite angles by considering the full Born-Infeld action.  Here, a
non-abelian version of the Born-Infeld action is necessary to make the
appropriate correspondence.  We analyzed the gauge theory spectrum
using Tseytlin's recent proposal for the symmetrized non-abelian
Born-Infeld action \cite{NDBI}.  Although we were unable to resolve
the discrepancy at finite angle in full, we found a strong hint in the
general structure of the action suggesting how this discrepancy might
be resolved.  It seems rather remarkable that the subtle
transcendental relations required to reproduce the correct spectrum of
strings attached to the D-branes are encoded in the symmetrized
Born-Infeld action.  Perhaps one could turn this type of argument
around and use the correspondence with string theory through T-duality
as a guiding principle for resolving various difficulties which plague
the non-abelian Born-Infeld action.

An intriguing aspect of the correspondence described here is related to
the spatial dependence of the fluctuations in the gauge theory and
string theory pictures for intersecting branes.  In the gauge theory
description, the off-diagonal fluctuations correspond to sections of
nontrivial $U(1)$ bundles over the torus, and are described in
\cite{vanBaal84} in terms of theta functions.
These theta functions were described in \cite{Hano} in terms of linear
superpositions along the lattice of harmonic oscillator
eigenfunctions.  This description essentially coincides with the
characterization of states in a constant magnetic field on the torus
in terms of Landau levels.  On the other hand, as argued by Berkooz,
Douglas and Leigh in \cite{BDL96}, the strings connecting intersecting
D-branes are essentially living in a harmonic oscillator potential
well and are localized to the intersection of the branes.  It is
natural to expect that the spatial dependence of the second quantized
string excitations in the intersecting D-brane picture will correspond
to gaussian and higher excited harmonic oscillator eigenfunctions.
This agrees nicely with the theta function description in the gauge
theory language; however, there is a subtlety in this relationship
because a T-duality transformation has been performed in some of the
directions on the torus.  Making this correspondence more precise
might lead to a better understanding of the action of T-duality as a
map between the Hilbert spaces of quantized Born-Infeld theory and
second quantized string theory.

Although in this paper we have concentrated on stationary backgrounds,
the configurations of tilted and intersecting branes which we have
considered here are closely related to boosted branes and systems of
branes scattering from one another.  For these systems, a convenient
dynamical quantity to compute is the phase shift due to scattering
\cite{bachas,Lifschytz:1996a}.  A discrepancy in the
relativistic corrections to D-brane scattering processes as computed
in a quantum
mechanics model derived from Yang-Mills theory was discussed in
\cite{vijayFinn97}.  This discrepancy seems similar in nature to the
$\theta/2$ vs. $\tan (\theta/2)$ discrepancy we have described here.
The authors of \cite{vijayFinn97} also consider
the possibility of resolving this discrepancy by considering the full
Born-Infeld action, but encounter similar difficulties in making the
exact correspondence at subleading order in backgrounds.  In
particular, they report a discrepancy between the phase shift computed
from the relativistic 0-brane quantum mechanics (equation (63) of
\cite{vijayFinn97})
$$
2 \delta_{00}^{QM}(v) \sim \left(\frac{v}{1-v^2} \right)^3
$$
and from the semiclassical phase shift in the Eikonal approximation
(equation (64) of \cite{vijayFinn97})
$$
2 \delta_{00}^{SC}(v) \sim \frac{(1-\sqrt{1-v^2})^2}{v \sqrt{1-v^2}}.
$$
It is interesting to note that $2\delta_{00}^{QM}$ and
$2\delta_{00}^{SC}$ are related by a relation similar to
(\ref{relation})
$$ 2 \delta_{00}^{SC}(V) = \frac{1}{V}  \int_0^V dv\, (1-v)^{3/2}  \delta_{00}^{QM}(v).$$
Although it is not completely clear why the integral prescription
(\ref{relation}) would apply to the phase shift in resolving the
discrepancy between $2 \delta_{00}^{QM}$ and $2 \delta_{00}^{SC}$,
this observation provides an indication that by analyzing
the combinatorics of Tseytlin's symmetrized trace prescription
\cite{NDBI}, relativistic quantum mechanics of 0-branes might be made
to agree with the semiclassical calculation.

Another aspect of the D-brane/gauge theory correspondence which we
have not pursued here, but which is an interesting direction for
further study, is the relationship between the moduli spaces of vacua in the
two pictures and the corresponding 0-modes of the system.  It was
pointed out by van Baal in \cite{vanBaal84} that the 
physical 0-modes in the gauge theory picture can be determined
explicitly, and that their number corresponds to the expected
dimension of the moduli space of instantons.  It has been argued
\cite{witinst,doug,vafa} that the moduli space of vacua for the
intersecting D-brane system is precisely given by the appropriate 
instanton moduli space.  The approach we have discussed here may
provide a concrete framework in which this connection can be explored
in more detail.

The web of relations between string theories and gauge theories has
become an active area of investigation in light of our increased
understanding of the role D-branes play in the non-perturbative
dynamics of string theory and in reformulating string theory itself
though M(atrix) theory.  Backgrounds with constant fluxes have played
an important role in recent studies of extended objects in M(atrix)
theory \cite{BFSS,GRT:1996,BSS96,BC97}.  Although there is clearly more
remaining to be said regarding the exact definition of the non-abelian
Born-Infeld theory, it seems that when this theory is properly defined
it should provide a mechanism for calculating some rather nontrivial
results for systems of interacting D-branes within a field theory
context. 

\section*{Acknowledgements}

We thank V.\ Balasubramanian, C.\ Callan, O.\ Ganor, I.\ Klebanov, G.\
Lifschytz, S.\ Ramgoolam, A.\ Tseytlin, and P.\ van Baal for useful
discussions. The work of AH is supported in part by DOE grant
DE-FG02-91ER40671, the NSF Presidential Young Investigator Award
PHY-9157482 and the James S. McDonnell Foundation grant No. 91-48. The
work of WT is supported in part by the National Science Foundation
(NSF) under contract PHY96-00258.


\begin{thebibliography}{10}

\bibitem{polchinski}
J.\ Polchinski, ``Dirichlet-Branes and Ramond-Ramond Charges,'' \PRL {\bf 75}
  (1995) 4724.

\bibitem{Strominger:1996}
A.\ Strominger and C.\ Vafa ``Microscopic Origin of the Bekenstein-Hawking
  Entropy,'' \PL {\bf B379} (1996) 99-104, {\tt hep-th/9601029}.

\bibitem{cm}
C.\ Callan and J.\ Maldacena, ``D-brane Approach to Black Hole Quantum
  Mechanics,'' \NP {\bf B472} (1996) 591-610, {\tt hep-th/9602043}.

\bibitem{ed}
E.\ Witten, ``Bound States of Strings and p-Branes,'' \NP {\bf B460} (1996)
  335, {\tt hep-th/9510135}.

\bibitem{BFSS}
T.\ Banks, W.\ Fischler, S.\ H.\ Shenker, and L.\ Susskind, ``M Theory as a
  Matrix Model: A Conjecture,'' {\tt hep-th/9610043}.

\bibitem{li}
M.\ Li, ``Boundary States of D-Branes and Dy Strings,'' \NP {\bf B460} (1996)
  351.\ {\tt hep-th/9510161}.

\bibitem{ck}
C.\ G.\ Callan and I.\ R.\ Klebanov, ``D-Brane Boundary State Dynamics,'' \NP
  {\bf B465} (1996) 473-486, {\tt hep-th/9511173}.

\bibitem{Polchinski:TASI}
J.\ Polchinski, ``TASI Lectures on D-branes,'' NSF-ITP-96-145, {\tt
  hep-th/9611050}.

\bibitem{BDL96}
M.\ Berkooz, M.\ R.\ Douglas, and R.G.\ Leigh, ``Branes Intersecting at
  Angles,'' \NP {\bf B480} (1996) 265-278, {\tt hep-th/9606139}.

\bibitem{SanjayZack}
Z.\ Guralnik and S.\ Ramgoolam, ``Torons and D-Brane Bound States,'' {\tt
  hep-th/9702099}.

\bibitem{vijayrob97}
V.\ Balasubramanian and R.\ G.\ Leigh, ``D-Branes, Moduli and Supersymmetry,''
  {\tt hep-th/9611165}.

\bibitem{Breckenridge:1997}
J.\ C.\ Breckenridge, G.\ Michaud, and R.\ C.\ Myers, ``New Angles on
  D-branes,'' {\tt hep-th/9703041}.

\bibitem{Gauntlett:1997}
J.\ P.\ Gauntlett, G.\ W.\ Gibbons, G.\ Papadopoulos, and P.\ K.\ Townsend,
  ``Hyper-Kahler Manifolds and Multiply Intersecting Branes, {\tt
  hep-th/9702202}.

\bibitem{Behrndt:1997}
K.\ Behrndt and M.\ Cveti\v{c}, ``BPS Saturated Bound States of Tilted p-Branes
  in Type II String Theory,'' {\tt hep-th/9702205}.

\bibitem{Costa:1997}
M.\ S.\ Costa and M. Cveti\v{c}, ``Nonthreshold D-Brane Bound States and Black
  Holes with Nonzero Entropy,'' {\tt hep-th/9703204}.

\bibitem{Hambli:1997}
N.\ Hambli, ``Comments on Dirichlet Branes at Angles,'' {\tt hep-th/9703179}.

\bibitem{vanBaal84}
P.\ van~Baal, ``SU(N) Yang-Mills Solutions with Constant Field Strength on
  $T^4$,'' \CMP {\bf 94} (1984) 397-419.

\bibitem{NDBI}
A.\ A.\ Tseytlin, ``On Non-Abelian Generalisation of Born-Infeld Action in
  String Theory,'' {\tt hep-th/9701125}.

\bibitem{DBI}
R.\ G.\ Leigh, \MPL {\bf A4} (1989) 2767.

\bibitem{Taylor:1996}
W. Taylor, ``D-brane Field Theory on Compact Spaces,'' {\tt hep-th/9611042}.

\bibitem{GRT:1996}
O.\ J.\ Ganor, S.\ Ramgoolam and W.\ Taylor, ``Branes, Fluxes and Duality in
  M(atrix)-Theory,'' {\tt hep-th/9611202}.

\bibitem{witinst}
E.\ Witten, ``Small Instantons in String Theory,'' \NP {\bf B460} (1996) 541.

\bibitem{doug}
M.\ Douglas, ``Branes within Branes,'' {\tt hep-th/9512077}.

\bibitem{tHooft81}
G.\ 't~Hooft, ``Some Twisted Selfdual Solutions for the Yang-Mills Equations on
  a Hypertorus,'' \CMP {\bf 81} (1981) 267.

\bibitem{Fradkin-Tseytlin}
E.\ S.\ Fradkin and A.\ A.\ Tseytlin, \PL {\bf B163} (1985) 123.

\bibitem{acny}
A.\ Abouelsaood, C.\ Callan, C.\ Nappi and S.\ Yost, \NP {\bf B280} (1987) 599.

\bibitem{Argyres-Nappi}
P.\ Argyres and C.\ Nappi, \NP {\bf B330} (1990) 151.

\bibitem{MaldaSuss96}
J.\ M.\ Maldacena and L.\ Susskind, ``D-Branes and Fat Black Holes,'' \NP {\bf
  B475} (1996) 679-690, {\tt hep-th/9604042}.

\bibitem{DasMathur96}
S.R.\ Das and S.D.\ Mathur, ``Excitations of D-Strings, Entropy, and Duality,''
  \PL {\bf B365} (1996) 79-86, {\tt hep-th/9601152}.

\bibitem{hashimoto96a}
A.\ Hashimoto, ``Perturbative Dynamics of Fractional Strings on Multiply Wound
  D-branes,'' {\tt hep-th/9610250}.

\bibitem{DMVV96}
R.\ Dijkgraaf, G.\ Moore, E.\ Verlinde, and H.\ Verlinde, ``Elliptic Genera of
  Symmetric Products and Second Quantized Strings,'' {\tt hep-th/9608096}.

\bibitem{DVV97}
R.\ Dijkgraaf, E.\ Verlinde, H.\ Verlinde, ``Matrix String Theory,'' {\tt
  hep-th/9703030}.

\bibitem{hashimoto96}
A.\ Hashimoto, ``Dynamics of Dirichlet-Neumann Open Strings on D-branes,'' {\tt
  hep-th/9608127}.

\bibitem{orbifold}
L.\ Dixon, D.\ Friedan, E.\ Martinec, and S.\ Shenker, \NP {\bf B282} (1987)
  13-73; S.\ Hamidi and C.\ Vafa, \NP {\bf B279} (1987) 465.

\bibitem{vanBaal82}
P.\ van~Baal, ``Some Results for SU(N) Gauge-Fields on Hypertorus,'' \CMP {\bf
  85} (1982) 529-547.

\bibitem{Hano}
J.\ Hano, in {\it Manifolds and Lie Groups}, (Birkhauser, 1981).

\bibitem{bankssusskind}
T.\ Banks and L.\ Susskind, ``Brane-antibrane forces'', {\tt hep-th/9511194}.

\bibitem{bachas}
C.\ Bachas, ``D-Brane Dynamics,'' \PL {\bf B374} (1996) 37-42, {\tt
  hep-th/9511043}.

\bibitem{Lifschytz:1996a}
G.\ Lifschytz, ``Comparing D-branes to Black-branes,'' {\tt hep-th/9604156}.

\bibitem{vijayFinn97}
V.\ Balasubramanian and F.\ Larsen, ``Relativistic Brane Scattering,'' {\tt
  hep-th/9703039}.

\bibitem{vafa}
C.\ Vafa, ``Gas of D-Branes and Hagedorn Density of BPS States,'' \NP {\bf
  B463} (1996) 415-419, {\tt hep-th/9511088}; ``Instantons on D-branes,'' \NP
  {\bf B463} (1996) 435-442, {\tt hep-th/9512078}.

\bibitem{BSS96}
T.\ Banks, N.\ Seiberg, and S.\ Shenker, ``Branes from Matrices,'' {\tt
  hep-th/9612157}.

\bibitem{BC97}
D.\ Berenstein and R. Corrado, ``M(atrix) theory in various dimensions,'' {\tt
  hep-th/9702108}.

\end{thebibliography}
\end{document}